\documentclass[11pt,a4paper]{article}
\usepackage[a4paper,left=3cm,right=2cm,top=2.5cm,bottom=2.5cm]{geometry}

\usepackage{booktabs} 
\usepackage{graphicx}
\usepackage{multirow}

\usepackage{rotating}
\usepackage{subfig}
\usepackage{url}
\usepackage{todonotes}
\usepackage{algorithm2e}
\usepackage{amsmath}

\usepackage{authblk}

\author[1]{Henri Casanova\thanks{henric@hawaii.edu}}
\author[2]{John Iacono\thanks{iacono@nyu.edu}}
\author[1]{Ben Karsin\thanks{karsin@hawaii.edu}}
\author[1]{Nodari Sitchinava\thanks{nodari@hawaii.edu}}
\author[3]{Volker Weichert\thanks{weichert@cs.uni-frankfurt.de}}
\affil[1]{University of Hawaii at M\=anoa, USA}
\affil[2]{New York University, USA}
\affil[3]{University of Frankfurt, Germany}

\title{An Efficient Multiway Mergesort for GPU Architectures}
\begin{document}
\maketitle

\begin{abstract}
Sorting is a primitive operation that is a building block for countless
algorithms.  As such, it is important to design sorting algorithms that
approach peak performance on a range of hardware architectures.  Graphics
Processing Units (GPUs) are particularly attractive architectures as they
provides massive parallelism and computing power.  However, the intricacies
of their compute and memory hierarchies make designing GPU-efficient
algorithms challenging.  In this work we present \emph{GPU Multiway Mergesort}
(MMS), a new GPU-efficient multiway mergesort algorithm.  MMS employs a new
partitioning technique that exposes the parallelism needed by modern GPU
architectures.  To the best of our knowledge, MMS is the first sorting
algorithm for the GPU that is asymptotically optimal in terms of global
memory accesses and that is completely free of shared memory bank
conflicts.

We realize an initial implementation of MMS, evaluate its performance on
three modern GPU architectures, and compare it to competitive implementations
available in state-of-the-art GPU libraries.  Despite these implementations
being highly optimized, MMS compares favorably, achieving 
performance improvements for most random inputs.  Furthermore, unlike MMS,
state-of-the-art algorithms are susceptible to bank conflicts.  We find
that for certain inputs that cause these algorithms to incur large numbers of bank
conflicts, MMS can achieve up to a 37.6\% speedup over its fastest
competitor.  Overall, even though its current implementation is not fully
optimized, due to its efficient use of the memory hierarchy, MMS
outperforms the fastest comparison-based sorting implementations available
to date.

\end{abstract}
%
%


\maketitle

\newpage
\pagestyle{plain}
\section{Introduction}
\label{sec:intro}

Sorting is a fundamental primitive operation. Consequently, much effort has been 
devoted to developing efficient algorithms and their implementations on a wide range
of hardware architectures, and in particular on the Graphics Processing Units (GPUs) that
have become mainstream for High Performance Computing.
A key challenge when designing GPU algorithms is exploiting the
memory hierarchy efficiently~\cite{GPU-book,overview2}.  It is well-known that certain
patterns when accessing data stored in \emph{global memory}
result in \emph{coalesced} memory accesses, leading to greatly increased
memory throughput~\cite{GMEM,optimize,gmem-bandwidth}.  
While many algorithms employ these
efficient access patterns, they fail to minimize the total number of accesses.
Ideal global memory access patterns allow blocks of $B$ elements to be retrieved
in a single accesses.  This pattern closely matches that of external disks, and
if we equate global memory accesses to input/output operations (I/Os), we can analyze our algorithm
using the Parallel External Memory (PEM) model~\cite{nodari:pem}.  Therefore, we can use the PEM model
to design I/O efficient algorithms for the GPU that achieve an optimally minimal number of global memory accesses.

\begin{figure}[th]
  \centering
  \includegraphics[width=0.5\textwidth]{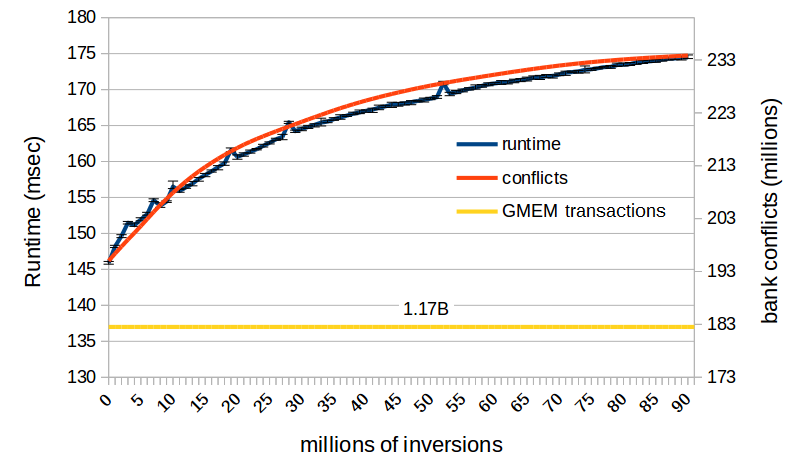}
  \caption{MGPU runtime (in ms) and number of bank conflicts vs. input sortedness when sorting $10^8$ 4-byte integers. Global memory transactions are shown to illustrate that they are independent of
input sortedness.  Global memory transactions do not correspond to either y-axis, though
they remain relatively constant at 1.17B and vary by at most 1\%.} 
  \label{fig:intro-conflicts}
\end{figure}

Access patterns are also important when
accessing the faster \emph{shared memory}: if multiple threads attempt to
access elements in the same \emph{shared memory bank}, a \emph{bank
conflict} occurs and accesses are then serialized.  Bank conflicts can lead
to significant performance degradation, which is often overlooked when
designing algorithms for GPUs~\cite{samplesort,green:thrust}.
To illustrate the impact of bank conflicts on performance,
we consider the current mergesort implementation in the modernGPU (MGPU) library~\cite{mgpu}.
This implementation employs a data-dependent merging operation in \emph{shared memory}.
When sorting a nearly-sorted
list, the access pattern results in very few bank conflicts.  A random
input sequence, however, has a randomized access pattern may result in more
bank conflicts.  Note that no other aspect of the MGPU mergesort implementation depends on the
input sequence (e.g., the number of global memory accesses depends only on the input \emph{size}).
Figure~\ref{fig:intro-conflicts} shows the runtime of MGPU mergesort, along with the
number of bank conflicts as reported by an execution profiler, when sorting
$10^8$ 4-byte integers on a K40 ``Kepler'' GPU~\cite{K40}.
Results shown are averaged over 10 trials on different input sets of varying
levels of ``sortedness.''  All input sets are created using a sorted list of
unique items and applying some number of inversions between random pairs.  The
x-axis corresponds to the number of such inversions, indicating the level
of sortedness of the input list.  
In Figure~\ref{fig:intro-conflicts} we see a clear correlation between the average
runtime of MGPU mergesort and the number of bank conflicts encountered.
Furthermore, we see that global memory transactions do not depend on input sortedness,
The close correlation between bank conflicts and runtime, along with the lack of other 
performance drivers that depend on input sortedness, leads us to conclude that bank conflicts
are a
performance driver.  Despite this potential performance loss due to
bank conflicts, however, MGPU mergesort remains among the best-performing
comparison-based sorting implementations available for GPUs~\cite{merry14:sortsurvey}.


In this work, we develop a new variant of the multiway mergesort (MMS)
algorithm for the GPU that (i)~is bank conflict-free, (ii)~achieves an asymptotically 
optimal number of global memory accesses, and (iii)~leverages the massive
parallelism available on modern GPUs.  To date, one can argue that there is
no true consensus regarding the use of theoretical models of computation
for designing efficient GPU algorithms. As a result, most GPU algorithms are
designed empirically.  By contrast, in this work we perform detailed
asymptotic analysis of I/O complexity using the 
PEM model.  Analysis shows that our MMS algorithm is more I/O-efficient 
than the standard pairwise mergesort approach used
in state-of-the-art GPU libraries.  We develop an initial implementation of
our algorithm, and show via experiments in three different GPU platforms
that our implementation is competitive with and often outperforms the fastest available
comparison-based sorting GPU implementations.  More
specifically, this work makes the following contributions:

\begin{itemize}

\item We propose the first, to the best of our knowledge, I/O efficient and
bank conflict-free (parallel multi-merge) sorting GPU algorithm;

\item We show via theoretical performance analysis that this algorithm
asymptotically outperforms the pairwise mergesort approaches currently used
in state-of-the-art GPU libraries;

\item We show experimentally that, when sorting large random input, an implementation of this
algorithm is competitive with highly optimized GPU libraries
(MGPU~\cite{mgpu} and Thrust~\cite{Thrust}) as well as a previously proposed
I/O-efficient parallel sample sort implementation~\cite{samplesort};

\item We also show experimentally that, because our algorithm's runtime
does not depend on the sortedness of the input due to it being bank
conflict free, for some input it can achieve a maximum speedup of
44.3\% and 37.6\% over two of the fastest available GPU libraries, respectively.

\end{itemize}

The rest of this paper is organized as follows.
Section~\ref{sec:background} provides background information on the PEM
model and GPUs. Section~\ref{sec:related} reviews related work.
Section~\ref{sec:multiway} describes our proposed algorithm.
Section~\ref{sec:analysis} provides comparative theoretical performance
analyses.  Section~\ref{sec:empirical} presents experimental results.
Section~\ref{sec:conclusion} concludes with a brief summary of results and
perspectives on future work.

\section{Background}
\label{sec:background}

In this section we first review the model we use to analyze
the I/O-complexity of our algorithm and its competitors, and then provide
relevant information on GPU architecture and the programming model.

\subsection{Parallel External Memory Model}
\label{sub:externalmemory}
The external memory model~\cite{aggarwal88:iomodel} is a well-known model analyzing the performance of algorithms 
that run on a computer with both a fast internal memory and a slow external memory, and
whose performance is bound by the latency for slow memory access, or I/O.  
The Parallel External Memory (PEM) model~\cite{nodari:pem} extends this model by
allowing multiple processors to accesses external memory in parallel.  In this work we use
the PEM model to design algorithms that efficiently use the global memory system of modern
GPU architectures.
The PEM model relies on the following problem/hardware-specific parameters:

\begin{itemize}
\item $N$: problem input size,
\item $P$: number of processors,
\item $M$: internal memory size, and
\item $B$: block size (the number of elements read or written during one external memory access).
\end{itemize}

In the PEM model, algorithm complexity is measured by the total number of I/Os.
For example, scanning an input in parallel would cost $O(\frac{N}{PB})$ I/Os.
Since this work is primarily concerned with sorting, we note that the lower bound on number
of I/Os to sort an input of size $N$ is~\cite{greiner:io}:

\begin{eqnarray*}
\text{sort}_{\text{PEM}}(N) = \Omega\left(\frac{N}{PB}\log_{\frac{M}{B}}{\frac{N}{B}} + \log{N}\right)
\end{eqnarray*}

\subsection{GPU Overview}
\label{sub:gpu}
\begin{figure}[th]
  \centering
  \includegraphics[width=0.5\textwidth]{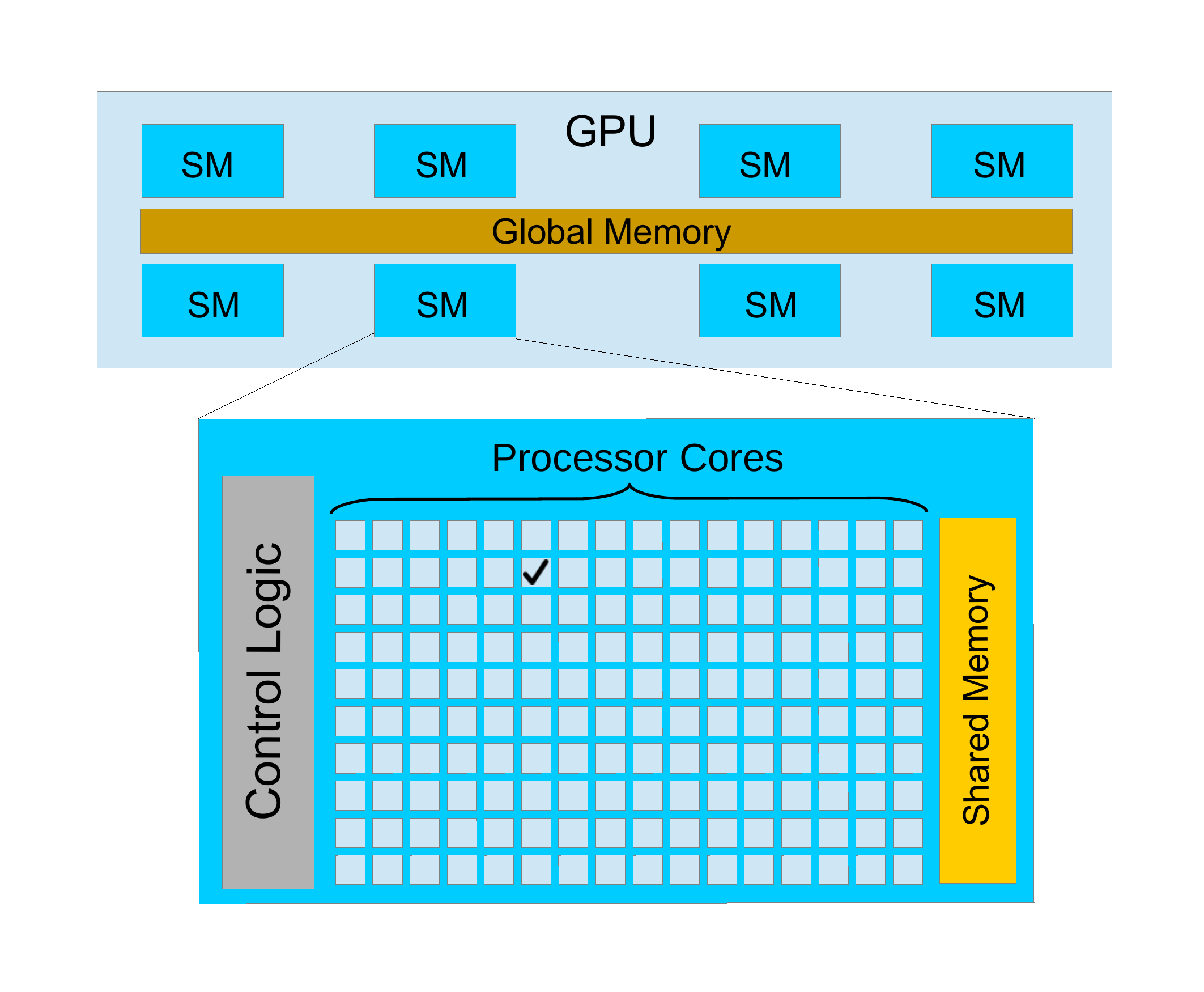}
  \caption{Illustration of a modern GPU architecture.}
  \label{fig:gpu-architecture}
\end{figure}

The PEM model was designed for multi-core systems with caches and RAM.
Modern GPU architectures, however, contain complex compute and memory
hierarchies (illustrated in Figure~\ref{fig:gpu-architecture}).  To design
efficient GPU algorithms, one must consider each level of these
hierarchies.

The compute hierarchy of most modern GPUs is designed to accommodate thousands of compute
threads~\cite{overview1}.  Physically, a GPU contains a number of \emph{streaming multiprocessors} (SMs), each of
which contains: instruction units, a shared memory cache, and hundreds of compute cores.
Logically, we consider the following computational units for GPUs:

\begin{itemize}
\item threads: single threads of execution,
\item warps: groups of $W$ threads that execute in SIMT (Single Instruction Multiple Threads) lockstep fashion~\cite{CUDA} ($W=32$ for most GPUs), and
\item thread blocks: groups of one or more warps that execute on the same SM.
\end{itemize}

In addition to the compute hierarchy, modern GPUs employ a user-controlled
memory hierarchy~\cite{overview2}.  Figure~\ref{fig:gpu-architecture} illustrates a
high-level view of the typical GPU memory hierarchy.

The largest component of the GPU memory hierarchy, \emph{global} memory,
is accessible by every thread executing on the GPU.  To achieve high memory 
throughput, all threads
within a \emph{warp} must access together consecutive elements in global memory.  This is called a
\emph{coalesced} memory access, allowing a warp to reference $W$ elements in a single access.
If we consider a warp
to be a single unit of execution, we say that
global memory accesses are performed in \emph{blocks}~\cite{CUDA}, where $B=W$ elements are read from a single access,
as with the PEM model discussed in Section~\ref{sub:externalmemory}.   Therefore,
we can apply the PEM model by equating global memory accesses as I/Os.

The smaller, fast \emph{shared} memory of the GPU is shared among threads within a thread block.  
All threads within a warp can each access shared memory locations in parallel. However, 
as mentioned in Section~\ref{sec:intro}, shared memory is organized in memory banks. 
If multiple threads within a warp attempt to access the \emph{same} memory bank, a \emph{bank conflict}
occurs and memory access are serialized~\cite{CUDA}.  Therefore, if all threads within a warp attempt
to access the same memory bank, a $W$-way bank conflict occurs, effectively reducing shared memory
throughput by a factor of $W$.

Finally, each thread has access to a set of fast private registers.  The
number of registers is limited and their access pattern must be known at
compile time.  Due to such limitations, it is difficult to model registers
for algorithm design and their utilization in CUDA programs can be viewed
as a low-level optimizations rather than a part of algorithm design.
Thus, we do not consider registers when designing algorithms,
although we use them in actual implementations.

\section{Related Work}
\label{sec:related}

Over the past decade, many works have focused on designing efficient
algorithms to solve a range of classical problems on the
GPU~\cite{dotsenko,merrill:scan,iptree,btree11,sengupta:scan,range,green:thrust}.
These works have introduced several optimization techniques, such as
coalesced memory accesses~\cite{GMEM,optimize,gmem-bandwidth}, branch
reduction~\cite{samplesort,smem-search}, and bank conflict
avoidance~\cite{smem-search,transpose}.  Several empirical models for
specific GPUs have been proposed that use
micro-benchmarking~\cite{owens11:model,hong14:model,bombieri16:model}, and
several fast GPU algorithms have been
produced~\cite{gmem-bandwidth,btree11,range} via the use of empirical
benchmarks~\cite{microbench} and the application of hardware-specific
optimization techniques to existing algorithms.  While all these approaches
can boost performance, abstract performance models are necessary to guide
the design of provably efficient GPU algorithms.  Several authors have
proposed such models, attempting to capture salient features of the compute
and memory hierarchies of modern GPUs in a way that balances accuracy and
simplicity~\cite{kothapalli09, hong09, ma14, DMM, HMM}.  Despite these
efforts, to date no model has been established as the definitive GPU
performance model.

Since the problem of sorting has been extensively studied over the past
half-century, in this section we focus on previous work relevant to sorting
on the
GPU~\cite{green:thrust,mgpu,CUB,samplesort,D-samplesort,merry14:sortsurvey,shearsort}.
According to a recent survey of several GPU
libraries~\cite{merry14:sortsurvey} the fastest currently-available sorting
implementations include the CUB~\cite{CUB}, modernGPU (MGPU)~\cite{mgpu},
and Thrust~\cite{Thrust} libraries.  CUB employs a GPU-optimized radix
sort, and thus can only be applied to primitive datatypes. MGPU and Thrust
use variations of mergesort (based on Green et al.~\cite{green:thrust}) and
many hardware-specific optimizations to achieve peak performance.  While
highly optimized, these mergesort implementations issue sub-optimal numbers
of global memory accesses and incur shared memory bank conflicts.
Leischner et al.~\cite{samplesort} introduced \emph{GPU samplesort}, a
distribution sort aimed at reducing the number of global memory accesses.
Their work was continued by Dehne et al.~\cite{D-samplesort} with a
deterministic version of the samplesort algorithm.  The work of Sitchinava
et al.~\cite{shearsort} focuses on shared memory only and presents an
algorithm that sorts small inputs in shared memory without bank conflicts.
Despite these efforts, no unified, provably efficient, and practical
sorting algorithm has been presented.  Thus, mergesort remains the
algorithm of choice in top-performing GPU libraries~\cite{Thrust,mgpu}.
The sorting algorithm introduced in this work illustrates an analytical
approach to designing GPU algorithms.  This algorithm minimizes global
memory accesses, incurs no shared memory bank conflicts, and outperforms
state-of-the-art implementations in practice.


\section{Multiway Mergesort}
\label{sec:multiway}
Mergesort is one of the most frequently used sorting algorithms today.  It is simple,
easily parallelizable~\cite{green14:mergepath} and load-balanced, and has optimal work 
complexity for comparison-based sorting.
However, most mergesort implementations rely on pairwise merging, resulting in $\log{N}$ merge
rounds to sort $N$ values.  In the context of the PEM model, described in 
Section~\ref{sub:externalmemory}, 
sequential pairwise merging requires $O(\frac{N}{B} \log{\frac{N}{B}})$ I/Os, and parallel mergesort
requires $O(\frac{N}{PB} \log{\frac{N}{B}})$ I/Os~\cite{nodari:pem}.  Note that this is a factor 
$\log{\frac{M}{B}}$ more memory accesses than optimal.  
An alternative that achieves the lower bound for I/O complexity is \emph{multiway mergesort}.

Like standard (pairwise) mergesort, multiway mergesort relies on repeated merge rounds,
however, at each
round, $K$ lists are merged into a single list (to achieve optimal I/O complexity, 
$K=\frac{M}{B}$).  To merge I/O efficiently, we must read and write only blocks of $B$ consecutive
elements and can only store a limited number of elements from each list in internal memory.  
Sequentially, this can be accomplished simply by using a minHeap and an output buffer.
Multiway merging in parallel, however, is difficult and may require increased
internal computation (which is ignored in the PEM model).   On modern GPUs,
both computation and memory accesses impact performance, so we develop a new method
of parallelizing multiway merging.

\subsection{Parallel Partitioning}
\label{sub:partition}
A large amount of parallelism is required to effectively use the computational power of modern GPUs.
Traditionally, multiway mergesort is not easily parallelized, so we introduce a technique
of partitioning the problem into independent tasks that can be executed in parallel.
We use the technique proposed by Hayashi et al.~\cite{hayashi:partition} to find the median among
$K$ lists (this is a generalization of the technique used by MGPU~\cite{green14:mergepath}).  The
technique is general and can be applied to find any $i$-th order statistic among $K$ lists.  This
method can be used to provide us with $P$ disjoint sets of elements (taken from all $K$ merge lists), $s_0,s_1,...,s_{p-1}$, 
where $P$ is the number of warps executing on the GPU.
Each of these sets, $s_i$, should have the property that its elements are \emph{non-overlapping} w.r.t
a given comparison function.  Formally, we say sets $s_i$ and $s_j$ are non-overlapping sets iff for all $a \in s_i$ and $b \in s_j$, $a \leq b$ implies $i \leq j$.

%

\begin{figure}[th]
  \centering
  \includegraphics[width=0.4\textwidth]{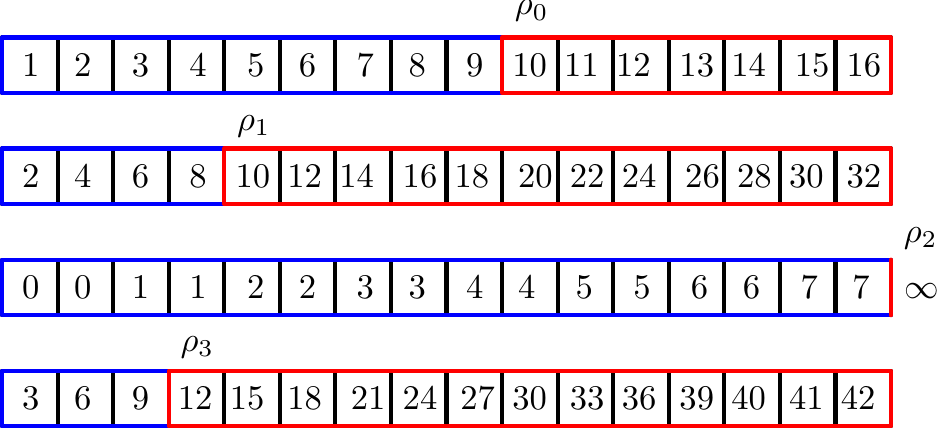}
  \caption{Example of a list of pivots creating 2 partitions.}
  \label{fig:pivots}
\end{figure}

Once each set $s_i$ is sorted, the concatenation of $s_0s_1...s_{p-1}$ will be sorted, as in quicksort.
Since elements in each $s_i$ are taken from the $K$ lists, sorting each set becomes
a smaller $K$-way merging problem.  We now have $p$ such smaller merging tasks, so each warp
can independently merge its own subset of the overall problem.
As described in the work of Hayashi et al.~\cite{hayashi:partition}, we find pivot indices in each of our
$K$ list by performing a binary search on each.  This can be accomplished in $O(K\log{N})$ time, where
$N$ is the length of each list and $K$ is the total number of lists.
Figure~\ref{fig:pivots} provides an example of such a set of pivots.

This process of partitioning can be executed in parallel for an arbitrary number of processors
(or warps), $P$.  Each warp performs a series of binary searches across $K$ lists to find its set of pivots
in $O(K \log{N})$ I/Os.  Each warp can then proceed with merging the values within its partition.



\subsection{I/O Efficient Merging}
\label{sub:iomerge}
Using the parallel partitioning method described above, we are able to obtain independent work
for each warp.  Therefore, we need only a warp-level
multiway merging algorithm to merge $K$ lists using $P$ warps.
To do this I/O efficiently, however, we must read and write blocks of $B$ consecutive 
elements at a time.  
As mentioned, this can be done sequentially using a heap and an output buffer, however
the GPU provides additional parallelism within the warp that we need to exploit.
Since $B$ consecutive elements must be read by a \emph{warp} at a time,
independent threads cannot work on separate sections and therefore must work in
parallel on the same data.
We do this using a new data structure called a \emph{minBlockHeap}.

A \emph{minBlockHeap} is a binary heap data structure where each node contains $B$ sorted values.
For any node $v$ with elements $v[0],\cdots,v[B-1]$, and child nodes $u$ and $w$, all elements contained in 
$u$ or $w$ must be greater than those contained in $v$ (i.e., $v[B-1] \le u[0]$ since each nodes
list is sorted). Therefore, our heap property If this
property is satisfied for our entire minBlockHeap, the root always contains the $B$ smallest elements.

We define the \emph{fillEmptyNode} operation that fills an empty node (i.e.,
a node without any elements in its list). Consider $v$ to be an empty node with non-empty
children $u$ and $w$.  W.l.o.g. assume that $u[B-1] > w[B-1]$.  The $fillEmptyNode(v)$ operation 
is performed as follows: merge the lists of $u$ and $w$,
fill $v$ with the $B$ smallest elements, fill $u$ with the $B$ largest elements, and set $w$ as empty.  
Since, prior to merging, $u$ had the largest element ($u[B-1]$), its new largest element has not changed
and the heap property holds for $u$.  We can continue down the tree by calling $fillEmptyNode(w)$ 
until we reach a leaf, which we can fill by loading $B$ new elements from memory.


To apply this data structure to multiway merging, we assign each of our $K$ input lists
to a leaf and build the minBlockHeap bottom-up using the \emph{fillEmptyNode} operation.
Each time a leaf node is empty, we fill it by reading $B$ new elements from the corresponding list
in global memory.
Once the minBlockHeap is built, we begin writing the root to global memory (in blocks of $B$)
as our sorted output.  We propagate the resulting empty node down to a leaf node and fill it from
global memory.  We repeat this process until we have merged all $K$ lists.  Note that
since we have $k$ leaf nodes, our minBlockHeap has a height of $\log{K}$ and a total of
$B(2K-1)$ total elements.  Thus if $K=\frac{M}{2B}$, our minBlockHeap will require 
$B(\frac{M}{B}-1) = M-B$ elements. 
The total number of I/Os of our multiway mergesort algorithm is
$O(\frac{N}{PB}\log_{\frac{M}{B}}{\frac{N}{B}})$, as we show in Section~\ref{sub:io}.

\subsection{Internal Memory Sort}
\label{sub:shearsort}

The multiway mergesort algorithm described in Section~\ref{sec:multiway} is a recursive
algorithm that sorts a list after $O(\log_{k}{N})$ merge rounds.  However,
if we employ an efficient internal memory sorting algorithm as a base case to the recursion, we can
reduce the number of merge rounds to $O(\log_{k}{\frac{N}{M}})$.  Since we will
be performing this sorting strictly in shared memory, we wish to avoid bank conflicts
(described in Section~\ref{sub:gpu}).  
We employ a variant of the \emph{shearsort} algorithm~\cite{sen:shearsort} that
efficiently uses GPU hardware and is bank conflict-free.  We note that, a work-efficient
variant of shearsort was introduced by Afshani et al.~\cite{peyman:shearsort} and may
be leveraged to improve performance.  However, we leave this optimization to future work.
The shearsort algorithm considers an input of $n$ values as a $\sqrt{n} \times \sqrt{n}$ matrix.
Shearsort sorts rows in alternating (ascending and descending) order, then columns in ascending order.
It repeats this process $\log{\sqrt{n}}$ times and, after a final sort of rows in ascending order,
the input is sorted. 

For our base case, we have each warp perform a shearsort on a $W \times W$ grid of elements,
where $W=32$ for most modern GPUs.  If we consider the grid of values to be in column-major order,
each row corresponds to one of our $W$ memory banks.  Therefore, each thread can sort a row
independently without any bank conflicts.  Sorting columns, however, will clearly result
in bank conflicts.  Since transposition can be performed efficiently without any bank
conflicts~\cite{transpose}, we transpose our matrix, allowing us to sort columns without any bank conflicts.
We repeat this process $\log{W}$ times to sort our base case of $W^2$ elements.  Each warp can
work on an independent set of elements, giving us a GPU efficient base case.
Note that we can extend the size of our base case by sharing data between pairs of warps
using bitonic merge.  This allows us to achieve an ideal base case size to minimize the 
number of merge rounds for a given $k$ and input size.  We discuss this and other
optimizations in more detail in Section~\ref{sub:tuning}.

\section{Performance Analysis}
\label{sec:analysis}
In this section, we discuss the asymptotic performance of our multiway merge algorithm as compared
to a standard pairwise merging technique.  As mentioned in Section~\ref{sub:externalmemory},
we consider performance in the context of the PEM model.  However, since modern GPUs employ
a hierarchy of memory and computation, we also look at the total work done by each algorithm
and the number of shared memory accesses and bank conflicts that may occur.

\subsection{I/O Complexity}
\label{sub:io}
In the context of the PEM model, we need only consider global memory accesses when measuring
algorithm complexity.  As discussed in Section~\ref{sub:gpu}, global memory is accessible by
all threads running on the GPU.  To achieve peak throughput, all threads in a warp must access
consecutive elements together.  When this occurs, we have a \emph{coalesced} memory access and
all $W$ threads receive their data in a single memory access.  This access pattern can be
seen as blocked access in the PEM model, so that a single I/O accesses $B$ elements (where $B=W$).
Therefore, we measure algorithm I/O complexity as the number of such accesses to global memory.

\subsubsection{Pairwise Mergesort}
The best-performing sorting algorithms available on the GPU today utilize pairwise mergesort.
In terms of I/Os, this is rather simple to analyze.  If we consider a total of $P$ warps running
concurrently on the GPU and an input of size $N$, the number of global memory accesses necessary
for pairwise mergesort defines the recurrence relation:

\begin{eqnarray*}
Q(N)  = \begin{cases}
  2Q(\frac{N}{2}) + O(\frac{N}{PB}), & \text{if $N>M$}. \\
  O(\frac{N}{P}), & \text{if $N\le M$}. 
\end{cases} \\
\end{eqnarray*}
\begin{eqnarray*}
 = O\left(\frac{N}{PB}\log_2{\frac{N}{M}}\right)
\end{eqnarray*}

This is the best we can hope to accomplish for pairwise mergesort, since we must access at least
$N$ elements during each merge round.  

\subsubsection{Multiway Mergesort}

The multiway mergesort algorithm introduced in Section~\ref{sec:multiway} aims to reduce the
I/O complexity by performing a $K$-way merge at each round.  At each merge round, each
warp performs a binary search in global memory to find its partition, as outlined in
Section~\ref{sub:partition}.  This results in $O(K\log{N})$ I/Os, therefore
the total I/O complexity for our multiway mergesort is: 

\begin{eqnarray*}
Q(N)  = \begin{cases}
  KQ(\frac{N}{K}) + O(\frac{N}{PB} + K\log{N}), & \text{if $N>M$}. \\
  O(\frac{N}{P}), & \text{if $N\le M$}. 
\end{cases} \\
\end{eqnarray*}
\begin{eqnarray*}
Q(N) = O\left( \frac{N}{PB}\left( \log_K{\frac{N}{M}}+1\right) + K\log{N}\log_K{\frac{N}{M}}\right)
\end{eqnarray*}

Assuming that $K$ is smaller than $\frac{N}{PB\log N}$, the cost of merging dominates, and our
asymptotic I/O complexity becomes:

\begin{eqnarray*}
Q(N) = O\left( \frac{N}{PB}\log_K{\frac{N}{M}}\right)
\end{eqnarray*}

Asymptotically, multiway mergesort has a factor $O(\log{K})$ less I/Os than a standard pairwise
mergesort.  To obtain optimal I/O complexity, $K=\frac{M}{B}$.  However, on modern GPU hardware it
is not clear how much internal memory we can assign to each warp while maintaining enough parallelism
for peak performance.  As discussed in Section~\ref{sub:gpu},
each SM has a fixed amount of shared memory, which can limit the number of warps running concurrently
on each SM.  Since GPUs depends on hyperthreading to hide
memory latency, the amount of memory per warp (i.e., $M$), and therefore $K$, must be carefully selected.  
In Section~\ref{sub:tuning} we empirically measure performance for a range of values for $K$.

\subsection{Internal Computation}
\label{sub:work}
Since algorithm runtime on modern GPUs may not depend on global memory accesses alone,
we also consider shared memory accesses during algorithm analysis.  As mentioned in
Section~\ref{sub:gpu}, the usability of registers is limited and we consider their use to
be an optimization technique.  Therefore, we can consider shared memory access to be
our smallest unit of work in internal computation.

\subsubsection{Pairwise Mergesort}
Rather than analyzing the details of a particular pairwise mergesort implementation
available in one of the GPU libraries, we consider the work complexity of a general
pairwise mergesort algorithm.  
We note that Thrust and MGPU, two of the
fastest mergesort implementations on the GPU, use a variation of the Merge Path~\cite{green14:mergepath}
technique to achieve parallelism.  Therefore, our analysis of a general mergesort algorithm
includes this technique as well.  
Since the number of shared memory bank conflicts for
most implementations are data dependent, we do not attempt to analytically model their impact
and instead measure the impact empirically in Section~\ref{sec:empirical}.  However,
we expect for an average input, bank conflicts will increase the number of shared memory
accesses needed by at least a factor of 2.

At each merge round, our general pairwise mergesort algorithm must: 1) find pivot points
for each \emph{warp} in global memory, 2) load portions of each list into shared memory and 
find pivots for each \emph{thread}, and 3) have each thread merge its own
section.  Thus, the total parallel internal computation time is:

\begin{eqnarray*}
T(N)  = \begin{cases}
  2T(\frac{N}{2}) + O\left(\frac{N}{PW} + \log{N} + \frac{N}{PM}\log{M}\right), & \text{if $N>M$}. \\
  O(N \log{N}), & \text{if $N\le M$}. 
\end{cases} \\
\end{eqnarray*}
\begin{eqnarray*}
 = O\left(\frac{N}{PW}\log_2{N}\right)
\end{eqnarray*}

Note that $PW$ is the total number of threads running on the GPU, making this algorithm
asymptotically optimal in terms of parallel internal computation. 

\subsubsection{Multiway Mergesort}
Unlike a typical pairwise mergesort algorithm, each step of our MMS algorithm
employs data structures internally.  While this may not increase the number of global memory
accesses (I/Os), it may increase the cost of internal computation.  Therefore, we analyze each
step of MMS to determine the total asymptotic parallel cost of internal computation.

At each merge round, each warp first finds $K$ pivots to define its partition.
This is done by $K\log{N}$ search steps, described in Section~\ref{sub:partition}.  
Once the partitions are found, each warp performs its $K$-way merge using the minBlockHeap
described in Section~\ref{sub:iomerge}.  Blocks of $B$ elements are read in to a leaf node, must pass
through the heap, and are outputted from the root.  Recall that each time we output a block of 
$B$ elements from the root, we call the \emph{fillEmptyNode} method on each of the
$\log{K}$ levels of the heap.  Each call to \emph{fillEmptyNode} involves merging two nodes
of $B$ elements each.  We do this in parallel using all $W=B$ threads of a warp using a
bitonic merging network in $2\log{W}$ time.  Thus, the total parallel internal computation
of each $K$-way merge is:

\begin{eqnarray*}
T(N)  = \begin{cases}
  KT(\frac{N}{K}) + O\left(\frac{N}{PW}\log{K}\log{W} + K\log{N}\right), & \text{if $N>M$}. \\
  O(N \log{N}), & \text{if $N\le M$}. 
\end{cases} \\
\end{eqnarray*}
\begin{eqnarray*}
 = O\left(\frac{N}{PW}\log_2{N}\log{W}\right)
\end{eqnarray*}

We see that MMS requires an additional factor $\log{W}$ internal
computation, compared with pairwise mergesort.  However, on modern GPU hardware, $W=32$, so
$\log{W}=5$.  Furthermore, MMS is bank conflict-free, while pairwise mergesort
implementations may incur up to $W$-way bank conflicts, resulting in a potential performance
loss of $O(W)$.  In Section~\ref{sec:empirical} we attempt to verify this hypothesis by
measuring the empirical performance of pairwise mergesort implementations.

\begin{table*}[t!]
\caption{Specifics of our three hardware platforms.}
\begin{tabular}{|l|l|l|c|c|c|c|}
\hline
Name & GPU Model & Architecture & Global memory & SMs & Cores per SM & Shared memory \\
\hline \hline
Gibson & GTX 770 & Kepler & 4GiB & 8 & 192 & 48 KiB \\
\hline
Algoparc & M4000 & Maxwell & 8GiB & 13 & 128 & 96 KiB \\
\hline
UHHPC & K40m & Kepler & 12GiB & 15 & 192 & 48 KiB \\
\hline
\end{tabular}
\label{tab:hardware}
\end{table*}

\section{Empirical Performance Results}
\label{sec:empirical}
The analysis in the previous section indicates that our MMS algorithm
provides key advantages over a pairwise mergesort algorithm.  In this
section, we evaluate the performance of our implementation of MMS on a
range of hardware platforms and input.  We measure execution time,
throughput (number of elements sorted per second), as well as numbers of
bank conflicts and numbers of global memory accesses.

\begin{figure*}[h]
    \centering
    ~ 
    \subfloat[]{
        \centering
        \includegraphics[width=0.45\textwidth]{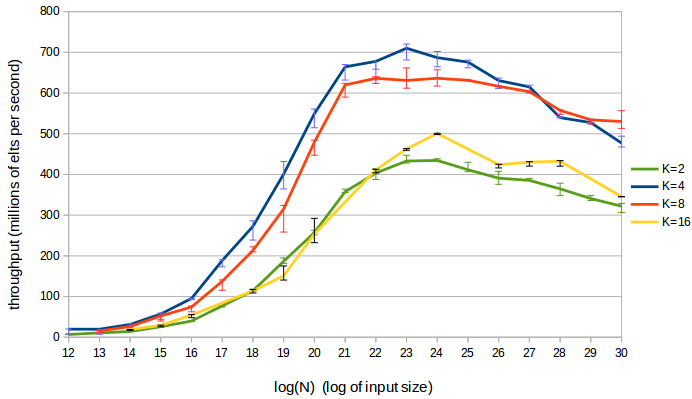}
}
    ~ 
    \subfloat[]{
        \centering
        \includegraphics[width=0.45\textwidth]{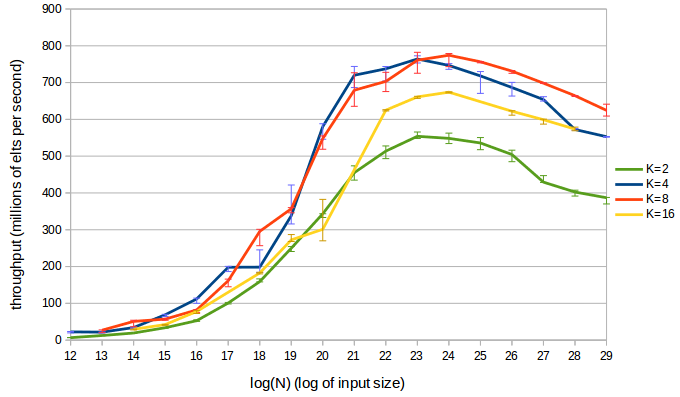}
}
    \caption{Average throughput vs. $N$ for $K=2,4,8,16$ on (a) UHHPC and (b) Algoparc.}
\label{fig:paramK}
\end{figure*}

\begin{figure}[th]
  \centering
  \includegraphics[width=0.5\textwidth]{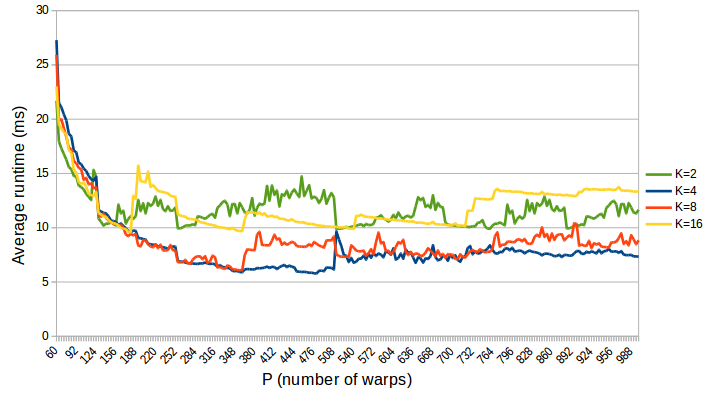}
  \caption{Average runtime of MMS vs. $P$ for $K = 2,4,8,16$ on UHHPC for $N=2^{22}$.  Error bars are omitted for readability.}
  \label{fig:paramP-cluster}
\end{figure}

\subsection{Methodology}
We consider three hardware platforms, each with a different modern graphics
card. All computations are performed on on the graphics cards, and no
attempt is made to use CPU compute resources.  Furthermore, execution times
are measured as time spent computing on the GPU, while time to transfer
data between the CPU and GPU is not included, as is customary in these
types of experiments.  The specifications of the GPUs of our three 
platforms are listed in Table~\ref{tab:hardware}.  On all platforms we use GCC
4.8.1 and CUDA 7.5, and all experiments are compiled with the -O3 optimization
flag.  Performance metrics such as bank conflicts are obtained via the nvprof profiling
tool~\cite{NSIGHT}, included in the CUDA 7.5 toolkit.  Since running the nvprof tool
impacts performance, execution time is measured on separate runs using the \emph{cudaEvent}
timer that is available with CUDA.
Each experiment is
repeated ten times, and we report on mean values, showing min-max error bars when
non-negligible.

We compare the performance of MMS with three leading GPU sorting libraries:
Thrust 1.8.1~\cite{Thrust}, modernGPU (MGPU) 2.10~\cite{mgpu}, and CUB
1.6.4~\cite{CUB}.  Thrust and MGPU implement pairwise mergesort algorithms and
provide the fastest comparison-based sorts available on the GPU.  CUB
provides the highest-performing radix sort.  Although CUB is not a
comparison-based sort, and is therefore limited to primitive
datatypes, we include it in some of our experiments for completeness.
We also include in some of our experiments the I/O-efficient samplesort implementation
in~\cite{samplesort}.


\subsection{Implementation Optimization and Parameter Tuning}
\label{sub:tuning}

\begin{figure*}[t]
    \centering
    ~ 
    \subfloat[]{
        \includegraphics[width=0.50\textwidth]{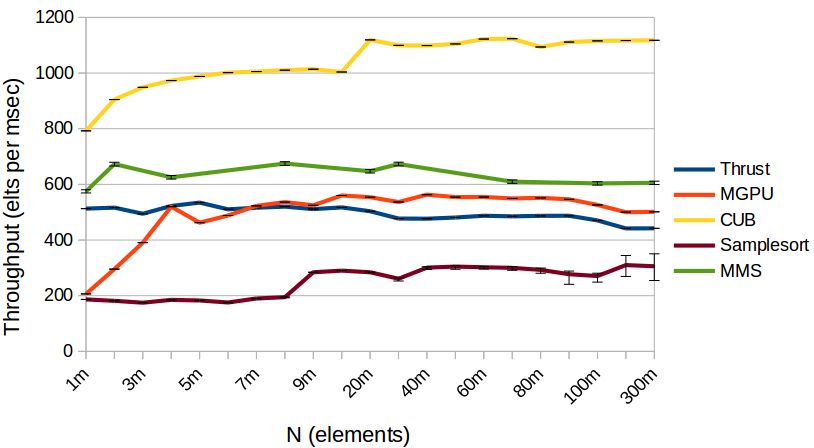}
    }
    ~ 
    \subfloat[]{
        \centering
        \includegraphics[width=0.50\textwidth]{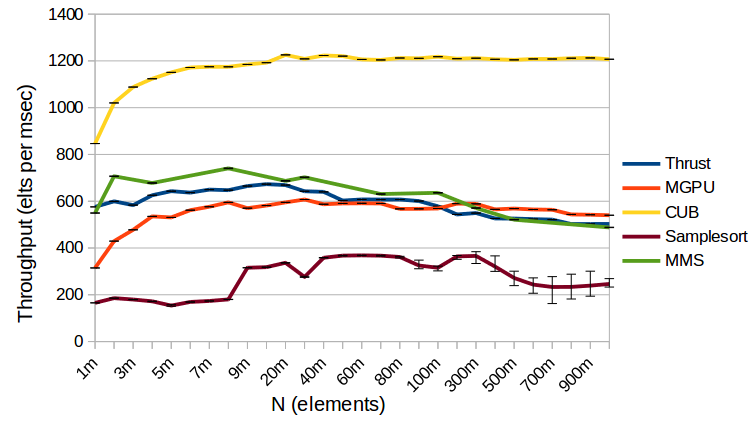}
}\\
    ~ 
    \subfloat[]{
        \centering
        \includegraphics[width=0.50\textwidth]{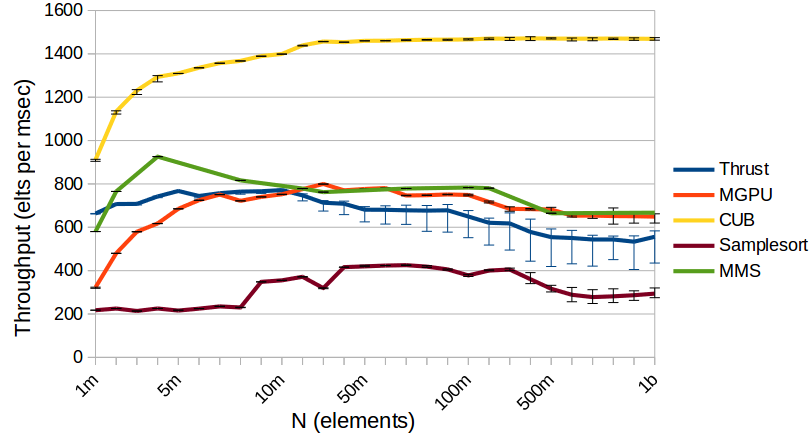}
}
    \caption{Average throughput vs. input size for fully random input
on (a) Gibson, (b) UHHPC, and (c) Algoparc.}
    \label{fig:rand-result}
\end{figure*}

We employ several typical optimization techniques in our implementation of
MMS~\footnote{ Note, however, that our implementation remains relatively
un-optimized compared to the MGPU and Thrust implementations.}.
First, when working independently threads make use of registers, which is
straightforward to implement and improves performance significantly.
Second, we use the warp shuffle~\cite{CUDA} operation that has been
available in nVidia GPUs since the Kepler architecture.  This operation
lets threads within a warp communicate between registers without relying on
shared memory. The shuffle operation leads to performance improvements when
the register access patterns of particular threads are deterministic.  As a
result, we use it to implement the bitonic merge used by the
minBlockHeap data structure (described in Section~\ref{sub:iomerge}), and
to avoid the transposition step of our internal memory shearsort (described
in Section~\ref{sub:shearsort}).  An third straightforward optimization is
to vary the base case size depending on $N$, so as to avoid an additional
merge round when input sizes do not precisely fit.  MMS performs
$\lceil\log_K{\frac{N}{|\text{basecase}|}}\rceil$ merge rounds.  For some
input sizes, one merge round will involve fewer than $K$ lists and work
will be wasted.  The larger $K$, the larger the performance loss due to
this wasted work.  Therefore, we simply increase the base case size so as
to reduce the number of necessary merge rounds.

Two parameters that have a key influence on the performance of MMS are $K$
(the number of lists merged at each round) and $P$ (the number of warps).
We determine good values for these parameters empirically for each 
platform.  We measure average execution time
over 10 trials for $K$ and $P$ value combinations ($K \in \{2,4,8,16\}$ and
$P \in [60,1000]$) when sorting an input of size $N=2^{22}$. For each $K$
value we then determine the empirically best $P$ value.
Figure~\ref{fig:paramP-cluster} shows results of this experiment on the
UHHPC platform.

On each platform, for a given $K$ value, we set $P$ to the best value
determined in the previous experiment, and measure average throughput as
$N$ varies.  We repeat this for each $K \in \{2, 4, 8, 16\}$.
Figure~\ref{fig:paramK} shows results of these experiments for the UHHPC
and Algoparc platforms.  We do not show results for the Gibson platform
because it is similar to UHHPC (both are ``Kepler'' generation GPUs), and
thus lead to the same conclusion that $K=4$ leads to the
best average performance (except for very large inputs).  However, for
Algoparc (which is a ``Maxwell'' generation GPU), $K=8$ is best for most
values of $N$.  The primary difference between the Maxwell and Kepler GPUs
is the amount of shared memory per SM.  Our Maxwell GPU has 96KiB of shared
memory per SM, while our Kepler GPUs have only 48KiB per SM.  The larger
$K$, the larger the amount of shared memory utilized by each warp. The size
of the shared memory thus limits the number of warps that can be running
concurrently on each SM, explaining why a larger shared memory allows for
the effective use of a larger $K$ value.  We note that, when $K=2$, our algorithm
becomes pairwise mergesort and has the same I/O complexity as the mergesort
algorithms used by other GPU libraries. As expected, when $K=2$, MMS achieves
significantly poorer performance than when $K=4$ or $K=8$.

\subsection{Experimental Results}
\label{sub:results}
In this section our main performance metric is the average throughput,
which we measure for each sorting algorithm implementation for various
input and input sizes.  As explained in Section~\ref{sec:intro}, we
generate input of varying ``sortedness,'' since sortedness has a large
impact on the performance of Thrust and MGPU.  By permuting each element in
our initial sorted list at least once, we generate a fully random,
uniformly distributed input (without repeats). Unless specified otherwise,
results are obtained by sorting input that consists of 4-byte integers.

\begin{figure}[th]
  \centering
  \includegraphics[width=0.5\textwidth]{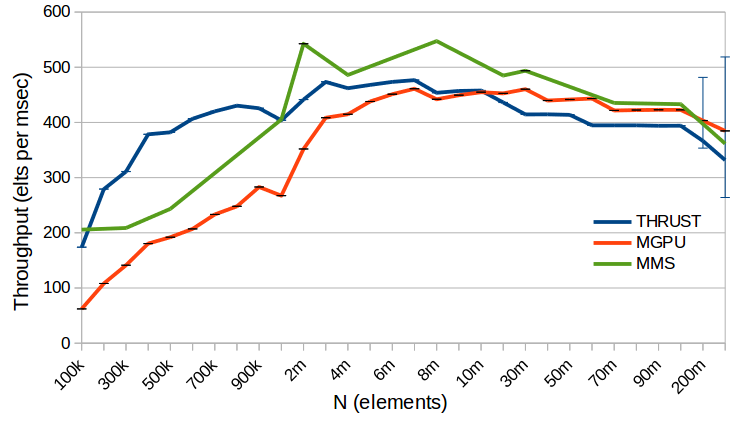}
  \caption{Average throughput vs. input size for full random inputs of 8-byte long datatypes
on UHHPC.}
  \label{fig:throughput-long}
\end{figure}

\begin{figure*}[ht]
    \centering
    ~ 
    \subfloat[]{
        \centering
        \includegraphics[width=0.45\textwidth]{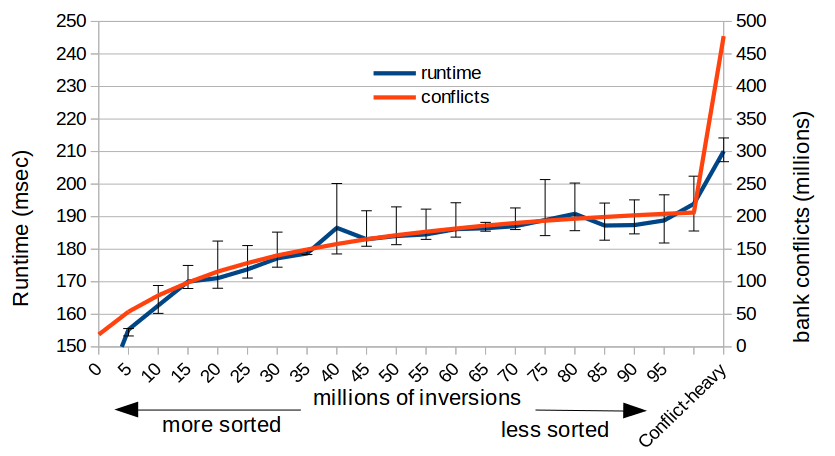}
}
    ~ 
    \subfloat[]{
        \centering
        \includegraphics[width=0.45\textwidth]{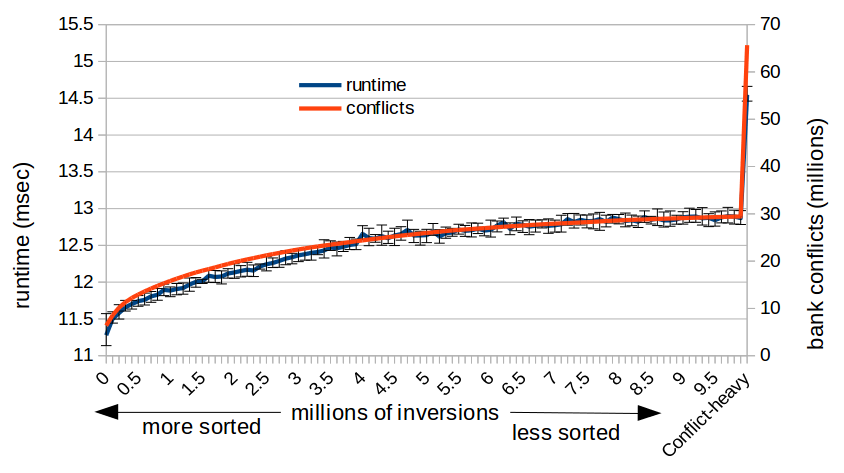}
}
    \caption{Average runtime vs. input sortedness for (a)~sorting $10^8$
    items using MGPU on the Gibson platform and (b)~sorting
    $10^7$ items using Thrust on the Algoparc platform. Results
    for conflict-heavy input are shown as the rightmost data point on the
    horizontal axes.}
\label{fig:conflict-runtime}
\end{figure*}
 
Figure~\ref{fig:rand-result} shows the throughput achieved by each
implementation when applied to fully random input of increasing sizes for
each of our three platforms.  These results show that MMS performs
comparably to the MGPU and Thrust comparison-based sorting implementations.
MMS outperforms both MGPU and Thrust for large input sizes.  This indicates
MMS provides better scalability, due to its optimal number of global
memory accesses and to the fact that it avoids all bank conflicts. As
expected, CUB, being a radix sort, achieves much higher throughput across
all input sizes. The samplesort implementation performs significantly worse
than all its competitors across the board.




Figure~\ref{fig:throughput-long} shows results on UHHPC for 8-byte integers
for MMS, MGPU, and Thrust (similar results are obtained on Gibson and
Algoparc). The trend in these results are similar to those observed for
4-byte integers, with MMS comparing favorably with its two competitors.

\subsection{Impact of Bank Conflicts}
\label{sub:worst}

One key feature of MMS is that it is shared memory bank conflict-free for
any input.  MGPU and Thrust, however, have memory access patterns that
depend on the input.  As seen in Section~\ref{sec:intro}, MGPU performs
increasingly worse as the input is more random (i.e., unsorted).  Since the
memory access patterns of MGPU and Thrust are deterministic, we should be
able to generate input that will cause these algorithms to incur large
numbers of bank conflict.  While such \emph{conflict-heavy} input may
result in performance degradation for Thrust and MGPU, the performance of
MMS, because bank conflicts are completely avoided, does not depend in the input.

To generate conflict-heavy input, we analyze the memory access pattern of
the mergesort algorithm in MGPU.  While Thrust employs a similar algorithm,
the code is more difficult to analyze. However, it turns out that
conflict-heavy generated based on our MGPU analysis is also conflict-heavy
for Thrust.  MGPU performs merging in shared memory by having each thread
merge two lists.  The lengths of these lists depend on the GPU generation.
On Kepler, resp. Maxwell, each thread merges lists of 11, resp. 15, items
at each round.  Note that these values are chosen to be co-prime with 32
(the number of banks) so as to reduce bank conflicts.  Each thread then
uses the Merge Path~\cite{green14:mergepath} method to find pivots and then
merges its items in shared memory.  Since the location of the elements
accessed by each thread depends on the input, we can generate
conflict-heavy input that result in large numbers of bank conflicts at
every memory access.  We create a small conflict-heavy input by hand and
generate larger conflict-heavy input by copying and interleaving
smaller conflict-heavy input. This methods generates input with 
numbers of items that are powers of two.

Figure~\ref{fig:conflict-runtime} shows average execution time and number
of bank conflicts vs. input sortedness, as defined in
Section~\ref{sec:intro}, for two sample experiments (MGPU on Gibson and
Thrust on Algoparc). The rightmost data point on the horizontal axis
corresponds to the conflict-heavy input generated as described above. These
results confirm that our conflict-heavy input does indeed lead to large
numbers of bank conflict for both MGPU and Thrust. In both cases it leads
to more than twice as many bank conflicts as when sorting fully random
input. These results further corroborate the preliminary results in
Figure~\ref{fig:intro-conflicts}: shared memory bank conflicts are one of
the key drivers of algorithm performance.

\begin{figure}[th]
  \centering
  \includegraphics[width=0.5\textwidth]{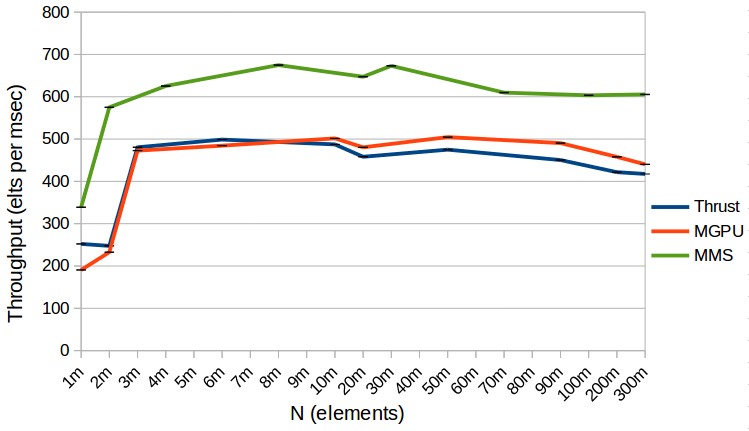}
  \caption{Average throughput vs.
           input size for \emph{conflict-heavy} input on the Gibson platform.}
  \label{fig:conflict-heavy}
\end{figure}


\begin{figure*}[ht]
    \centering
    ~ 
    \subfloat[]{
        \centering
        \includegraphics[width=0.45\textwidth]{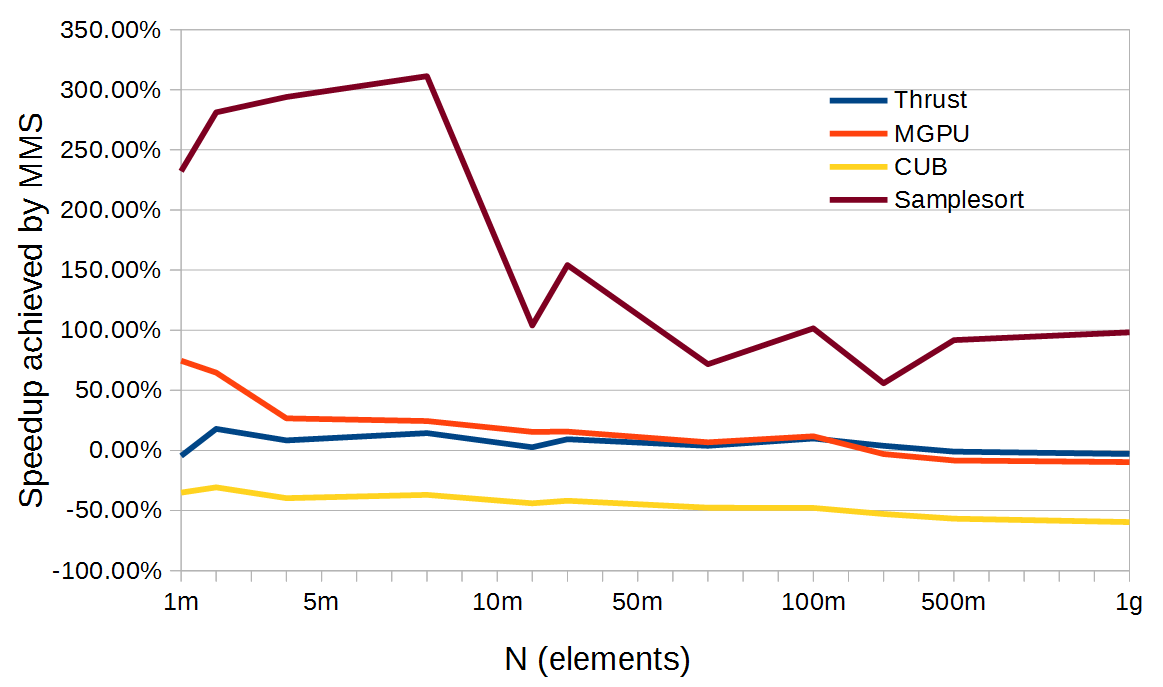}
}
    ~ 
    \subfloat[]{
        \centering
        \includegraphics[width=0.45\textwidth]{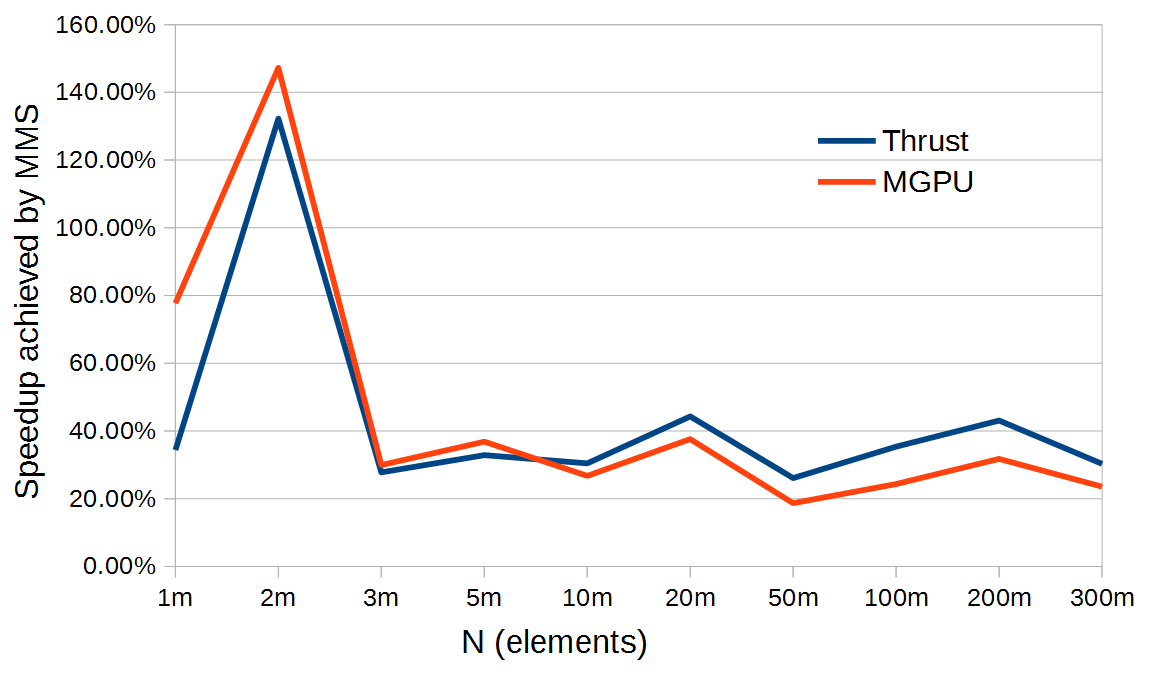}
}
\caption{Relative speedup of MMS over state-of-the-art sorting implementations.  Results
shown for \textbf{a)} random inputs on UHHPC platform and \textbf{b)} conflict-heavy inputs
on the Gibson platform.}
\label{fig:speedup}
\end{figure*}

Using our conflict-heavy input, we compare the average throughput of each
implementation.  Figure~\ref{fig:conflict-heavy} presents results on the
Gibson platform.  These results indicate that both Thrust and MGPU suffer from
significant performance degradation due to bank conflicts on the conflict-heavy input, while
MMS achieves performance similar to a random input sequence.  Figure~\ref{fig:speedup}
illustrates this performance degradation by showing the relative speedup
of MMS over existing algorithms for \textbf{a)} random and \textbf{b)} conflict-heavy inputs.  Ignoring
noise from small input sizes, Figure~\ref{fig:speedup}a indicates
that MMS is competitive with both Thrust and MGPU.  In Figure~\ref{fig:speedup}b, however,
we see that MMS achieves up to 44.3\% and 37.6\% speedup over Thrust and MGPU, respectively,
on the Gibson platform.  We
obtain similar results on our two other platforms.  On UHHPC, MMS
achieves a maximum speedup over MGPU and Thrust of 41.2\% and 36.8\%, respectively.  On
Algoparc MMS achieves
up to 31.9\% and 29.4\% speedup over MGPU and Thrust, respectively.


\subsection{Performance Bottlenecks for MMS}
\label{sub:bottleneck}

MGPU and Thrust provide highly optimized sorting implementations, and
their performance shortcomings are due to the nature of the sorting
algorithm itself rather than to overlooked implementation details. More
specifically, these implementations use a pairwise mergesort algorithm,
which does not achieve the I/O complexity lower bound and causes shared
memory bank conflicts. By contrast, MMS uses the memory hierarchy
efficiently, but could likely benefit from various optimizations in
addition to those mentioned in Section~\ref{sub:tuning}. In this section we
identify the bottleneck component in our current MMS implementation.

Recall from Section~\ref{sec:multiway} that the MMS algorithm consists of
a partitioning phase and a merging phase at each recursive level.  Using
the nvprof~\cite{NSIGHT} profiler, we find that, for the ideal $P$ values
determined as explained in Section~\ref{sub:tuning}, the partitioning phase
makes up less than 2\% of the overall execution time.  Furthermore, we find
that the node merging component of the \emph{fillEmptyNode} heap method
(described in Section~\ref{sub:iomerge}) contributes to more than 60\% of
the overall runtime.  This is due to the communication required between
threads.  More specifically, 32 threads (one warp) work on merging two
lists of 32 elements each, which is done using a bitonic merging network,
thus requiring a large amount of thread communication.   Although we leave
the optimization of this component of MMS for future work, its optimization
should greatly increase overall performance.

\section{Conclusions}
\label{sec:conclusion}

In this work we present MMS, a new GPU-efficient multiway mergesort
algorithm. By using the PEM model and considering shared memory access
patterns, we show that MMS achieves an optimally minimal number of global
memory accesses and does not cause any shared memory bank conflicts.
Furthermore, through the use of a new parallel partitioning method, MMS
exposes the high level of parallelism needed to approach peak GPU
performance.

We perform a detailed empirical analysis and compare performance results of
MMS with two highly optimized comparison-based sorting implementations,
MGPU and Thrust.  Our results show that MMS exhibits performance comparable
to MGPU and Thrust on randomly generated input, and outperforms them for
most large input.  A performance shortcoming of MGPU and Thrust is bank
conflicts, which we highlight by generating input that causes these
implementations to incur larger numbers of bank conflict.  On such input
MMS, whose performance is not sensitive to the input permutation, achieves 
up to 44.3\% and 37.6\% speedup over Thrust and MGPU, respectively.

Since MMS issues the optimal number of global memory accesses and avoids
bank conflicts, we expect it to scale more efficiently than standard
pairwise mergesort algorithms, such as MGPU and Thrust, as hardware
improves and memory capacities grow.  In addition, larger shared memories
will enable MMS to utilize a larger branching factor, further improving
its performance.


We leave several performance improvements to MMS as future work. Some of
the hardware-specific optimizations used by MGPU and Thrust may yield
performance gains for MMS as well.  Since merge operations on the
minBlockHeap structure are a significant bottleneck (see
Section~\ref{sub:bottleneck}), techniques such as pipelining should improve
MMS performance as well.  Finally, we may be able to develop a variation of
MMS that avoids the use of minBlockHeap altogether and instead relies on
the parallel partitioning technique to perform multiway merging.  Such an
algorithm may be able to avoid the additional internal work and other
drawbacks of using a heap, while retaining I/O efficiency.

\bibliographystyle{abbrv}
\bibliography{biblio.bib}

\end{document}